# Privacy-Enabled Biometric Search

Scott Streit
Computer Science Innovations, LLC
16487 Frederick Road
Woodbine, MD 21797
+1-301-596-2550
scott@scottstreit.com

Brian Streit
DRN
4150 International Plaza, Suite 800
Fort Worth, TX 76109
+1-301-706-0908
bqstreit@gmail.com

Stephen Suffian
Department of Electrical and
Computer Engineering
Villanova University
800 East Lancaster Avenue,
Villanova, PA 19085
ssuffian@villanova.edu

*Abstract*— Biometrics have a long-held hope of replacing passwords by establishing a non-repudiated identity and providing authentication with convenience. Convenience drives consumers toward biometrics-based access management solutions. Unlike passwords, biometrics cannot be script-injected; however, biometric data is considered highly sensitive due to its personal nature and unique association with users. Biometrics differ from passwords in that compromised passwords may be reset. Compromised biometrics offer no such relief. A compromised biometric offers unlimited risk in privacy (anyone can view the biometric) and authentication (anyone may use the biometric). Standards such as the Biometric Open Protocol Standard (BOPS) (IEEE 2410-2016) provide a detailed mechanism to authenticate biometrics based on pre-enrolled devices and a previous identity by storing the biometric in encrypted form. This paper describes a biometric-agnostic approach that addresses the privacy concerns of biometrics through the implementation of BOPS. Specifically, two novel concepts are introduced. First, a biometric is applied to a neural network to create a feature vector. This neural network alone can be used for one-to-one matching (authentication), but would require a search in linear time for the one-to-many case (identity lookup). The classifying algorithm described in this paper addresses this concern by producing normalized floating-point values for each feature vector. This allows authentication lookup to occur in up to polynomial time, allowing for search in encrypted biometric databases with speed, accuracy and privacy.

## I. INTRODUCTION

Convenience drives consumers toward the biometrics-based access management solutions. Studies show that 52% of smartphone users [5] want to use their fingerprints instead of a password, a further 61% want to use fingerprints to unlock their phones, and 48% want to use eye recognition. Biometrics have a long-held hope of replacing passwords by establishing a non-repudiated identity and providing authentication with convenience. While biometrics are increasingly becoming a preferred and convenient method for identity and authentication [4,6], there are currently serious concerns regarding the protection of biometric data. Biometrics include a wide range of information taken from a person, e.g., fingerprints, face, voice, iris pattern, etc. and his/her behavioral properties, e.g., gait, date, time, and location. Recent increases in the processing power and sensor technologies allow digital signal processing (DSP) algorithms to run in the time needed for a real-time authentication (1-5 seconds, similar to username and password login processing).

Unlike passwords, biometrics cannot be script-injected; however, biometric data is considered highly sensitive due to its personal nature and unique association with users. Biometrics differ from passwords in that compromised passwords may be reset. Compromised biometrics offer no such relief. A compromised biometric offers unlimited risk in privacy (anyone can view the biometric) and authentication (anyone may use the biometric). Since resetting the biometric is impossible, the privacy of biometrics is of utmost concern.

Consumer privacy concerns have been exacerbated by recent security breaches, such as those at Ashley Madison and Home Depot. For biometrics to penetrate the field of authentication and identification, users must remove themselves as impediments.

Current transport-level encryption technology provides strong protection of the transmission of all types of data, including biometric data, that satisfies confidentiality, assurance, and non-repudiation requirements. These methods protect from an adversary listening in on communication. However, current biometric techniques require the use of plaintext search for matching, which means the biometric is visible. This paper describes techniques for matching and searching that address this problem by conducting matching in the encrypted space, thus allowing for biometric data to be immediately encrypted upon collection.

## II. EASE OF USE

Encryption has been widely used for several millennia to protect information during transport [8]. The nature of the information dictated the levels and type of encryption required to protect it in transit in case of compromise [9].

The most popular form of modern encryption during data transit is the use of transport layer security (TLS) which is commonly called by its predecessor technology, secure socket layers (SSL). We generalize and abbreviate as (TLS/SSL). Additionally, 2-way SSL sits on top of TLS /SSL and is an identity protocol that uses a certificate exchange to identify an individual.

In the case of stored data, it has not been possible to encrypt all stored [15] data at the application level because of requirements for searching the data. Personally Identifiable Information (PII) requires encryption mechanisms and additional policies and processes [10,11,15] for data protection because any operation on the data demands decryption of that data for viewing and editing. For example,

HIPAA [15] requires the encryption of data in transport and offers policies for the release and dissemination of that data. The policy for cryptographic strength in this case is meant to protect against compromise of PII databases in cases of their theft. If operations could be performed, such as search, without decryption, data need not be exposed to potential compromise.

Biometric data requires further protections by processes and policy that introduce the need for additional mechanisms including more sophisticated encryption [19]. One-to-one record match given a reference biometric and a new input record has been solved [19] but encrypted match and encrypted search remain an open problem.

The current state of biometrics is as follows:

1. An enrollment takes place for a new device. This links a biometric to an identity. A key exchange takes place that links the enrolling device to an identity and a server. The biometric is stored either on the server or on the client.

2. Each instance of a biometric differs. The key exchange from step 1 offers an initial layer of security. A likeliness value is then calculated between the initial biometric and the new biometric.

3. A match result is then derived from step 2.

Currently, biometric searches require that the biometric be stored in a persistent and plaintext form. Existing stores are not general purpose and require the participants to agree to store their biometrics. It has thus far been infeasible, from a performance perspective, to search biometric data, even when it is stored as ciphertext, because the match is not direct. Two biometric instances of the same person vary, when in encrypted form, processing time becomes unacceptable as identification requires decryption of the biometric, and still an exhaustive search. Lastly and in some ways more importantly, the biometric is not one-way encrypted, meaning that ownership of the decryption key can allow for access to the plaintext biometric data.

*A. Biometric Open Protocol Standard (BOPS)*

In 2013, no standard existed for the interoperability of biometrics or a set of best practices for biometrics. In July of 2013, a set of researchers formed IEEE Working Group 2410, the Biometric Open Protocol. The initial standard described a set of web service interfaces and a set of inter-operable processing rules. Products were built using IEEE 2410-2015.

In late 2014, it became clear that there were impediments prohibiting the large scale dissemination of biometrics. Vulnerabilities that were discovered in the initial BOPS specification, such as client-side malware attacks, led to the revision of BOPS in 2016 (BOPS II). BOPS II considers privacy and the threat of malware taking over a client device while simultaneously considering the risks associated with plaintext storage on the server. It provides a detailed mechanism to authenticate based on pre-enrolled devices and a previous identity by storing the biometric in encrypted form.

BOPS II allows the match to take place in plaintext or as homomorphic encryption. The base practice is homomorphic encryption.

BOPS I and BOPS II are based on the tested computer-based implementation of the Trusted Computer System Evaluation Criteria (TCSEC). BOPS conforms to the TCSEC, which is the United States Department of Defense standard that sets basic requirements for assessing the effectiveness of computer security controls built into a computer system. TCSEC was created by the National Computer Security Center, an arm of the National Security Agency (NSA) and is also frequently referred to as "Orange Book, Section B1." BOPS also conforms to the Director of the Central Intelligence Directive 6/3 protection level 3, level 4, and level 5 (PL3, PL4, and PL5), and to the standards of the Multiple Independent Levels of Security/Safety (MILS) architecture.

There are two primary use cases in which BOPS II could be implemented. The first is the one-to-one case: sending and receiving encrypted biometric data and comparing to an existing encrypted biometric sample. We consider this the authentication use case because given a biometric vector and an identity as input, we can authenticate (or deny) a given biometric vector if it matches an existing biometric vector corresponding to that identity.

BOPS II also specifies functions for the one-to-many case: searching encrypted biometric records given a new biometric record. The only solution to this case is an $O(n)$ exhaustive search in which each record is decrypted and compared. We consider this the identity use case because given a biometric vector as input, we can search and determine whether or not the given biometric is in a database. When storing profiles of users, we wish to use a biometric to find a user. However, even if we abate the concern for privacy, two biometrics from the same person will not match exactly. The inexact nature of this match leads a non-polynomial search, which is untenable and not implementable.

Biometric identification in BOPS II may occur by storing the biometric in plaintext or the substantially more difficult method of storing the biometric data using homomorphic encryption, allowing for matching to be conducted in the encrypted space using Euclidean measurement algorithms. The risk to privacy is inversely proportional to the amount of time and operations that occur in the plaintext space. If biometrics are kept in plaintext one hundred percent of the time, this leaves a large window for any sort of privacy attack. Alternatively, if a biometric is gathered as ciphertext and all manipulations are done as ciphertext, there is no place where the biometric is plaintext and vulnerable. Matching in the encrypted space using a one-way encryption offers the highest level of privacy.

BOPS II provides for data privacy in biometric store, biometric match and biometric search. Clearly, data privacy requires the encryption of the biometrics. The question is whether a solution exists that allows us to work entirely in the encrypted space. BOPS II as a specification offers a general framework for biometric operations, but it does not offer a specific means of implementation. This paper provides a solution compliant with BOPS II for privacy-based match and search in a biometric-agnostic way, allowing for biometric matching in the encrypted space.

The rest of the paper is organized as follows. In Section 3, we formulate the problem with conducting biometric matching and search in encrypted space. In Section 4, we discuss our Methodology, which includes designing a Neural Network to acquire a feature vector that is then sent through a novel classification algorithm, allowing for biometric search between log(n) and n time.

## III. PROBLEM FORMULATION

The issue with current biometric schemes is they require a mechanism for: (1) acquiring the biometric, (2) plaintext biometric match, (3) encrypting the biometric, (4) performing a Euclidean-measurable match, and (5) searching using a one-to-many indexing scheme.

To do steps 1 through 5 for every biometric is time consuming, error-prone and frequently nearly impossible to do before the biometric becomes deprecated. Our goal is to develop a scheme, techniques and technologies that allow us to work with biometrics in a privacy-protected and polynomial-time based way that is also biometric-agnostic. We will use machine learning to solve problems (2) - (5). We focus much of our effort on (5) and allow points (1) - (4) to provide supplemental information.

Finally, complete content and organizational editing before formatting. Please take note of the following items when proofreading spelling and grammar:

### A. Encryption

Since we have little control over devices such as cameras or sensors, the biometric template arrives as plaintext. If we encrypt it immediately and only process it as ciphertext, we have the maximum practical level of privacy. An important part of offering this highest level of privacy is a one-way encryption algorithm, meaning that given ciphertext, there is no mechanism to get to the original plaintext. Many one-way encryption algorithms exist, such as MD5 and SHA-512. However, these algorithms are not homomorphic. This means we cannot do a closeness match between two ciphertext vectors using Euclidean measurements.

This paper explores and produces a general purpose solution that produces biometric ciphertext that is Euclidean-measurable. We do this using a Neural Network. We then apply a classification algorithm to allow for one-to-many identification. This solution maximizes privacy and runs between $O(1)$ and $O(\log(n))$ time.

## IV. METHODOLOGY

In this Section, we first design a Neural Network for the encryption of a biometric, and then use a classification algorithm on the encrypted output to perform both one-to-one authentication and one-to-many identification. Matrix multiplications and several other operations are used in the neural network to translate an Initial Biometric Vector (IBV) into a Euclidean Measurable Feature Vector (EMFV), which is then stored. When a new biometric is submitted for matching/identification, this Current Biometric Vector (CBV) is sent through the neural network so it can be translated into an EMFV. These EMFVs are then used in a separate classification algorithm to perform one-to-many search between $O(1)$ and $O(\log(n))$ time.

This paper uses the biometric of facial recognition to test the approach. However, the method was designed specifically to be implementable regardless of the type of biometric. We believe similar techniques will work for voice, fingerprints and a host of other biometrics. This is due to the use of feature vectors for encryption, which are the output of a neural network. The proposed classification algorithm acts on these feature vectors. This allows this general approach to be used despite varying biometrics, neural networks, and data.

### A. Our Defined Neural Network

We use a convolutional neural network for processing images and determining the correct cost function. The spatial size of the output volume can be computed as a function of the input volume size $W$, the kernel field size of the convolutional layer neurons $K$, the stride with which they are applied $S$, and the amount of zero padding $P$ used on the border. The formula for calculating how many neurons "fit" in a given volume is given by $1 + (W - K + 2P)/S$. One of the objectives of this research is to find the ideal number of neurons and layers in the neural network. Fig. 1 below shows an activation function as a summation of a matrix of weights times the $X$ values. This yields a $Y$ value.

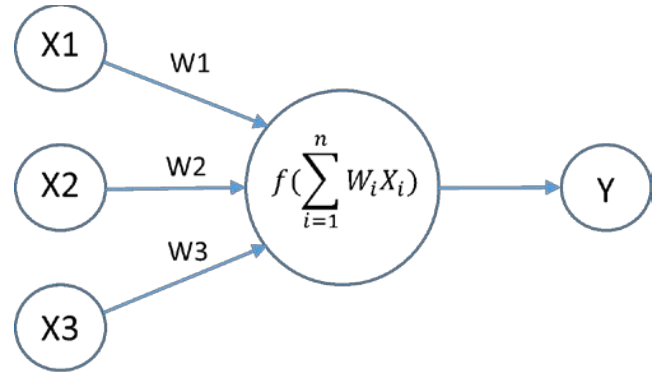

Figure 1. Example of Activation Function

The neural network is initially trained as a classifier using labeled biometric data. However, the traditional use of a neural network in this regard is neither novel nor scalable as it assumes the person is already stored and the images, from the training, are present. One cannot assume, or even fathom that we can train on every person available to our set.

Instead, when an IBV is introduced into the neural network, the vector used in the n-1 layer can be used as a unique feature vector to identify that IBV but in the homomorphic encrypted space. This feature vector is Euclidean-measurable and encrypted.

The feature vector replaces the biometric vector. The feature vector is a list of 256 floating-point numbers. It is impossible to reconstruct the biometric from a list of 256 floating-point numbers. The feature vector is therefore a one-way encryption. Feature vectors have the property of being Euclidean-measurable.

Appendix 1 shows an example of our neural network. The initial image is applied to the neural network. We see the resulting softmax, which shows which of the defined "buckets" the image belongs. In this bucket, we have the images for the given individual. Next it shows us the matrix computations to create the softmax. Two convolutional steps prior to the softmax is fc7-conv which has the output from convolutional layer 7. This output is the feature vector.

For example, if the feature vector for person $X$, as 256 values, when applied to the classifying function results in a floating-point number of $Y$. For all values near $Y$, we can identify the individual. We can only do this if our feature vector is Euclidean-measurable and classifying function is stable.

Appendix 2, shows us the entire network. We have a variety of convolutional layers together with a rectifier (ReLU) and pooling nodes. The rectifier is, as of the publication of this paper, the most popular activation function for deep neural networks. The rectifier is an activation function defined as $f(x) = \max(0, x)$.

Beyond ReLU, our convolutional network uses pooling which is a form of non-linear down-sampling. We use a type of pooling algorithm called maxpooling, which partitions input images into a set of non-overlapping rectangles and, for each such sub-region, outputs the maximum. Once a feature has been found, its exact location is not as important as its relation to other features. The function of the pooling layer is to progressively reduce the spatial size of the representation to reduce the amount of parameters and computation in the network. Through the use of convolutions, ReLU and maxpooling, we get a reasonable size vector of 256 to use for matching, classifying and comparison.

The resultant encrypted, Euclidean-measurable feature vectors (EMFVs) from the neural network can later be used in the classification algorithm to receive a scalar value that can be compared to the feature vectors from other biometrics.

*B. Our Classification Algorithm*

To use the feature vector for storage, we need to index it. For indexing we use a classification algorithm that, given an input feature vector, returns a floating-point number. This floating-point number is an index for storage. The classification algorithm has a dependence on a high-quality mean vector, which aids in creating distances between persons. The classification algorithm allows us to search and find the enrolled person in polynomial time.

The aforementioned floating-point number allows us to index search for a previously stored biometric. An appropriate classification algorithm gives us the intended person based on the input feature vector. Given a feature vector of 256 nonnegative integers, the procedure given below classifies this vector into the appropriate pool. Each pool is for one and only one distinct person. If the person is not found, it is assumed the algorithm will look for neighboring persons to complete the search.

The idea is to use distances to classify unknown vectors of the form $x = (x_0, x_1, \cdots, x_{255})$. Unless otherwise noted, $x_i$ is a nonnegative integer for all $i \in \{0,1,\cdots,255\}$. We define the distance between two vectors as $d(x,y) = \sum_{i=0}^{255}|x_i - y_i|$. Let $U$ be the set of known vectors and $|U|$ the number of vectors in $U$. We will assume $|U| < \infty$.

The mean of all vectors in $U$ is given by $m_U = 1/|U|\sum_{x \in U} x$ and can be calculated explicitly if $|U|$ is sufficiently small. If $|U|$ is not sufficiently small, then $m_U$ can be approximated. Note that the coordinates of $m_U$ are not necessarily nonnegative integers.

Consider a partition of $U$ as $U = \coprod_j P_j$. We have observed experimentally that for each $P_j$, there exist $a_j, b_j \in (0, \infty)$ such that $d(m_U, y) \in [a_j, b_j]$ for all $y \in P_j$. Moreover, $[a_j, b_j]$ is disjoint from $[a_k, b_k]$ for $j \neq k$. In other words, bands of distances from the mean vector can be used to classify unknown vectors.

Given a vector $y \notin U$, we calculate $d(m_U, y)$ to determine how to uniquely extend partitions of $U$ to partitions of $V = U \cup \{y\}$. If $d(m_U, y) \notin [a_j, b_j]$ for all $j$, then the interval $[a_j, b_j]$ closest to $y$ is chosen and the subset $P_j$ associated with this interval is chosen to include $y$ in the extension of the original partition to a partition of $V$. If it happens that $y$ is equidistant to two different intervals, then the subset in which to include $y$ in the partition of $V$ is not well-defined. In this case, the numerical results should be re-examined and a better choice for at least one of the two intervals equidistant to $y$ should be made.

V. RESULTS

We used 257 images for training. The Neural Network is a 32 node, 8 convolution layer network. The training allocates the appropriate weights to the network. With this trained network, we apply a new face for either storage or lookup. In either case, we put the face through the neural network and at convolutional layer 7 we have a feature vector.

Using a set of images, and applying our neural network we receive the following EMFV at convolutional layer 7. An example of part of one such output is below:

0.83215457, 0, 0.97926098, 2.34792662, 6.1617651, 6.31998634 2.36983943, 3.691468, 3.82195878, 0.94054723, 1.48430574, 3.68506813, 3.71155834, 0.73430908, 1.13436747, 2.62325954, 2.40342426, 1.69623852, 1.55817235, 1.41169822, 0.89341599

For a set of 15 persons and 257 images the results are desirable. The feature vector classifies to a specific person within runtime $O(1)$. Using the algorithm, as specified in Section 4.1, we have ranges of three individuals. Using normalized vectors, the resulting ranges of images are as follows. For Person 1, our normalized distances range from .85 to 1.12. For Person 2 our range is from 1.18 to 1.32. For Person 3, our normalized distances range from .39 to .68.

As a practical example, subsequent IBVs of person one, when given to the neural network and classification algorithm, will produce results between .85 and 1.12. Similarly, subsequent IBVs for person 3 will yield results between .39 and .68. These acceptable ranges offer us banding for each person. Therefore, the idea is we see banding and no collisions for the bands between persons. For small sets of data, our results are exact.

## VI. DISCUSSION

Our hypothesis is as follows:

1. We can use deep learning to do more than categorize.

2. The feature vectors from machine learning are Euclidean-measurable and stable.

3. We use feature vectors for one to one match guaranteeing us homomorphic encryption. We have the highest level of privacy.

4. Our classification algorithm allows us to search, given a feature vector as input.

Conclusions 1 through 3 appear exact. Conclusion 4 requires more data and testing.

The numeric ranges after the application of our classification algorithm offer us buckets for search. Each bucket results in the selection of an individual.

## VII. FUTURE WORK

We anticipate continued work in the area of indexing based on feature vectors. The results hold for smaller sets, we wish to see if the indexing and classification algorithm holds over set of hundreds of thousands of individuals.

We will also experiment with the mean vectors looking for wider banding between the categorization values on a person by person basis.

We also will do a combinatorial analysis to determine the order of the classification algorithm. We will find if the runtime of the classifier is $O(1)$ or $O(\log(n))$.

## VIII. CONCLUSION

In conclusion, we see that deep learning solves the homomorphic encryption problem for one to one match. Adding our classification algorithm, to the results of the neural network offers us a solution for search.

We match based upon the IBV as input, the application of the neural network the resulting feature vector. We then use the feature vector, with our classification algorithm to provide indexed search.